\documentclass[runningheads]{llncs}
\usepackage[T1]{fontenc}
\usepackage{graphicx,verbatim}
\usepackage{amsmath}
\usepackage{amssymb}
\usepackage{cite}
\usepackage{multirow}
\usepackage{booktabs}
\usepackage[ruled,vlined]{algorithm2e}
\begin{document}
\title{Denoising via Repainting: an image denoising method using layer wise medical image repainting}
%

\author{Arghya Pal, Sailaja Rajanala, CheeMing Ting, Raphael Phan}  
\authorrunning{Anonymized Author et al.}
\institute{Monash University \\
    \email{arghya.pal@monash.edu}}

\maketitle              
\begin{abstract}
Medical image denoising is essential for improving the reliability of clinical diagnosis and guiding subsequent image-based tasks. In this paper, we propose a multi-scale approach that integrates anisotropic Gaussian filtering with progressive Bezier-path redrawing. Our method constructs a scale-space pyramid to mitigate noise while preserving critical structural details. Starting at the coarsest scale, we segment partially denoised images into coherent components and redraw each using a parametric Bezier path with representative color. Through iterative refinements at finer scales, small and intricate structures are accurately reconstructed, while large homogeneous regions remain robustly smoothed. We employ both mean square error and self-intersection constraints to maintain shape coherence during path optimization. Empirical results on multiple MRI datasets demonstrate consistent improvements in PSNR and SSIM over competing methods. This coarse-to-fine framework offers a robust, data-efficient solution for cross-domain denoising, reinforcing its potential clinical utility and versatility. Future work extends this technique to three-dimensional data.

\end{abstract}

\section{Introduction}
\label{sec_intro}
Denoising is a fundamental challenge in MRI, with applications ranging from surgical planning to diagnostic interpretation. While conventional approaches (e.g., wavelet thresholding, nonlocal means) and unsupervised or self-supervised strategies (e.g., Aniso~\cite{perona1990scale}, nonlocal~\cite{guidotti2009new}, Beltrami~\cite{ben2023denoising} have shown promise for reducing high-frequency noise, they often struggle to maintain fine structures and subtle details critical for clinical decision-making. Recent advances in deep learning DnCNN~\cite{7839189}, N2N~\cite{pmlr-v80-lehtinen18a}, Ne2Ne~\cite{Huang_2021_CVPR}) have demonstrated significant improvements in reconstruction quality by learning end-to-end mappings between noisy inputs and clean targets. However, these learned models typically require extensive labeled data and may not generalize well to unseen noise distributions or different imaging modalities.

In this work, we propose a novel denoising framework that integrates \emph{multi-scale anisotropic filtering} and a \emph{progressive component-redrawing} strategy to robustly mitigate noise in diverse medical images. As illustrated in Fig.~\ref{fig:paths} (top), our method first constructs an \emph{anisotropic Gaussian blur pyramid} (i.e., a scale space) to incrementally isolate noise from salient anatomical features. Beginning with a heavily smoothed representation and progressing to finer scales, we adaptively preserve edges, corners, and textural patterns across modalities. We then segment each partially denoised image into coherent regions and redraw these regions using parametric Bézier paths, thereby maintaining the structural integrity of the underlying tissue boundaries. This coarse-to-fine redrawing strategy ensures that large homogeneous regions are captured early, while smaller or more complex structures are refined in subsequent iterations.

Compared to purely data-driven solutions, our method does not require extensive labeled training data, as it inherently incorporates spatial continuity constraints through anisotropic smoothing. Moreover, the proposed formulation accommodates multiple noise levels without additional retraining, making it especially suitable for real-world clinical scenarios where noise characteristics can vary widely. Experimental results on multiple MRI datasets, including AxFLAIR brain MRI and Cor-PD knee MRI, demonstrate that our approach consistently achieves state-of-the-art performance in terms of both PSNR and SSIM, while also retaining critical anatomic details. By effectively balancing noise reduction and structure preservation, the presented framework broadens the applicability of denoising methods to a wide range of medical imaging applications.

\section{Methodology}
\label{sec_method}

\begin{figure}[!t]
    \centering
    \includegraphics[width=\linewidth]{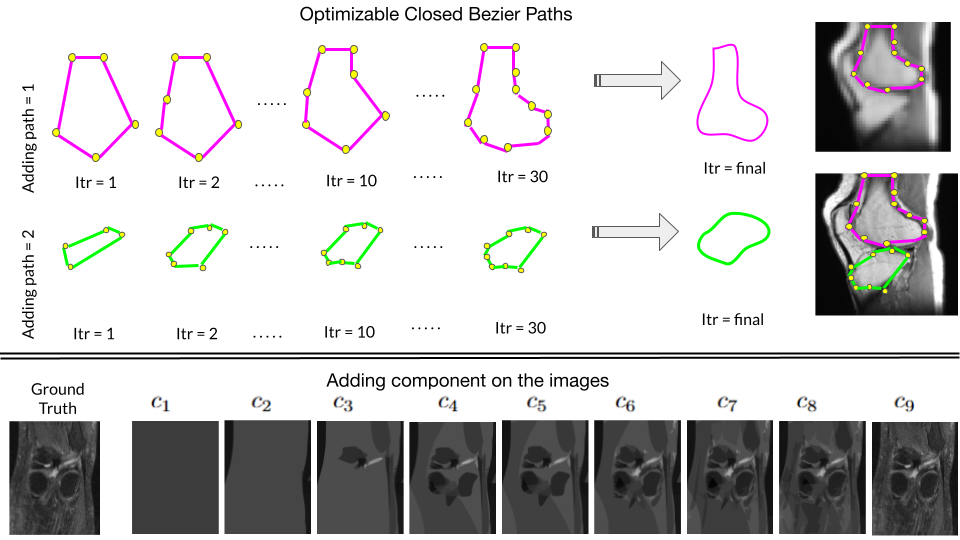}
    \caption{(\textbf{Top}) Our methodology (\S\ref{sec_method}) progressively adds Bezier paths for each component in a coarse-to-fine manner. (\textbf{Bottom}) Demonstration of denoising an image in a layer-wise coarse-to-fine manner with redrawing using a small number of paths. $c_1,\cdots,c_9$ indicate the path numbers.}
    \label{fig:paths}
\end{figure}

Our proposed approach is designed to denoise a given image by leveraging a multi-scale, anisotropic Gaussian blur pyramid (i.e., a scale space), followed by progressive component identification and redrawing. As shown in Fig.~\ref{fig:paths} (top), the denoising is carried out in a coarse-to-fine manner by incrementally adding Bezier paths that represent image components with coherent color and texture. We detail each step below.

\subsection{Step 1: Scale-Space Pyramid}
\label{sec:scale_space}
We begin with the assumption that no prior knowledge of the noise distribution is available. Hence, the core objective of this step is to isolate essential image structures (e.g., edges, corners) from unstructured high-frequency noise. To achieve this, we construct an \emph{anisotropic Gaussian blur} scale-space pyramid, motivated by the principles in~\cite{perona1990scale}.

\paragraph{Anisotropic Gaussian Blur.} Let $I$ be the original noisy image. We define the $t$-th scale-space image $I_{t}$ by:
\begin{equation}
    I_{t} = G(\mu_t, \sigma_t) * I \;+\; \nabla I \;+\; \Delta I,
    \quad t = 0, \ldots, T,
\end{equation}
where $G(\mu_t, \sigma_t)$ is an anisotropic Gaussian kernel parameterized by \(\mu_t\) and \(\sigma_t\). By varying these parameters from large (producing a strongly blurred image) to small (retaining finer details), we obtain multiple scales of smoothing. The operators \(\nabla I\) and \(\Delta I\) represent the gradient and Laplacian of the noisy image \(I\), respectively.

\begin{figure}[!t]
    \centering
    \includegraphics[width=\linewidth]{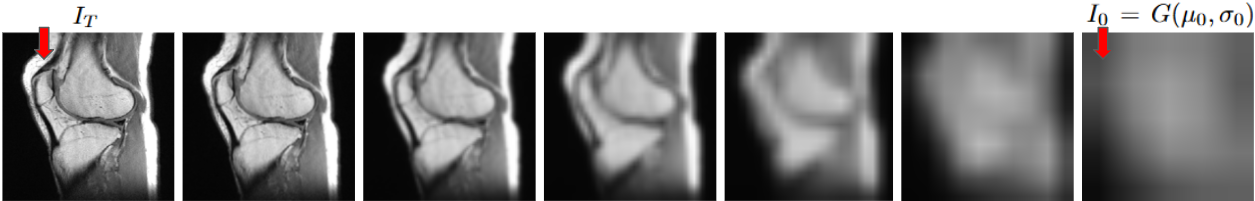}
    \caption{\emph{Anisotropic Gaussian Blur} with varying $G(\mu_t, \sigma_t)$~\cite{perona1990scale}. Larger values of \(\mu_t\) and \(\sigma_t\) yield stronger smoothing, while smaller values preserve finer details.}
    \label{fig:scalespace}
\end{figure}

\paragraph{Scale-Space Construction.} We form a sequence of images \(\{I_{0}, I_{1}, \ldots, I_{T}\}\), each providing a different trade-off between noise suppression and edge preservation:
\begin{itemize}
    \item Larger \(\sigma_t\) and \(\mu_t\) values favor stronger smoothing to eliminate high-frequency noise.
    \item Smaller \(\sigma_t\) and \(\mu_t\) values preserve detailed structures (e.g., edges, corners).
\end{itemize}
Figure~\ref{fig:scalespace} illustrates the impact of varying $(\mu_t, \sigma_t)$. In practice, \(I_{0}\) (the first level) is often the most heavily smoothed image and serves as the basis for subsequent redrawing.

\subsection{Step 2: Progressive Component Redrawing in $I_{0}$}
\label{sec:redrawing}
Once we obtain \(I_{0}\), a partially denoised and smoothed image, we segment it into connected components \(\{ c_{1}, \ldots, c_{N} \}\) based on color or texture continuity. The assumption is that spatially contiguous pixels with similar color attributes can be grouped into coherent regions, each corresponding to a uniform color/texture area.

\paragraph{Bezier-Path Representation.} For each component \(c_n\), we redraw its shape and color using a parameterized Bezier path (Fig.~\ref{fig:paths}, top). 
We represent each component as a sequence of \(k\) cubic Bézier segments, where each segment \(i\) is defined by control points \(\{\mathbf{P}_{i,j}\}_{j=0}^k\) and parameterized for \(j \in [0,1]\) by
\[
\mathbf{B}_i(t) \;=\; (1 - t)^3 \,\mathbf{P}_{i,0}
\;+\; 3t(1 - t)^2 \,\mathbf{P}_{i,1}
\;+\; 3t^2(1 - t)\,\mathbf{P}_{i,2}
\;+\; t^3\,\mathbf{P}_{i,3}.
\]
To ensure these segments form a closed loop, we impose the positional continuity constraints
\[
\mathbf{P}_{i,3} \;=\; \mathbf{P}_{(i+1)\bmod n,\,0}
\quad\text{for}\quad i=0,\dots,n-1,
\] 
This path captures the topological outline of the component and assigns a representative color derived from \(I_{0}\). Following~\cite{xu2022live}, we employ an \(l_1\)-based color difference metric to check the fidelity of the redrawn region compared to the partially denoised image:
\[
   \text{Diff}(c_n, I_{0}) = \|\,c_n^{(\text{redrawn})} - c_n^{(I_{0})}\|_{1}.
\]
If this discrepancy is below a predefined threshold (e.g., 0.1), the component is accepted. Otherwise, the Bezier path is refined, or a new initialization is considered.

\subsection{Step 3: Iteration \& Optimization Across Scales}
\label{sec:iter_optim}
We refine the denoising at multiple resolutions by repeating the redrawing procedure for the subsequent images \(I_{1}, I_{2}, \ldots, I_{T}\) in the scale-space pyramid. Specifically:
\begin{enumerate}
    \item \textbf{Identify New Components at Scale $I_{t}$.} Additional or refined structures not visible at $I_{t-1}$ emerge, denoted \(\{ c_{N+1}, \ldots, c_{N+M} \}\).
    \item \textbf{Add/Refine Bezier Paths.} We introduce new Bezier paths for newly identified components \(\{ c_{N+1}, \ldots, c_{N+M} \}\) and may increase control points in previously identified paths to improve boundary accuracy.
    \item \textbf{Update Reconstructed Image.} By overlaying each newly rendered or refined component, the image is incrementally “painted” in a coarse-to-fine manner.
\end{enumerate}

Figure~\ref{fig:paths} (bottom) showcases how each new component addition helps to fill in critical details while mitigating noise. Larger regions (e.g., background) are captured first at coarser scales, while finer structures (e.g., edges, tissue boundaries) are accurately detailed at higher-resolution scales. This iterative strategy balances noise reduction with preserving essential image topology and color continuity. 
We represent the methodology in Algorithm 1. 
\begin{algorithm}[!h]
\caption{Proposed Multi-Scale Anisotropic Rendering}
\label{alg:concise}
\DontPrintSemicolon
\KwIn{
    Noisy image \(I\), \\
    Number of scale-space levels \(T\), \\
    Learning rates \(\alpha, \beta\), \\
    Regularization weight \(\lambda\), \\
    Iterations per scale level \(t_{\max}\)
}
\KwOut{Denoised Image}
\BlankLine

\textbf{Step 1: Build Scale-Space Pyramid (Sec.~\ref{sec:scale_space})}\;
Construct \(\{I_{0}, \dots, I_{T}\}\) via anisotropic Gaussian blur at multiple scales.

\BlankLine
\textbf{Step 2: Initialization at Coarsest Scale (Sec.~\ref{sec:redrawing})}\;
Identify initial components in \(I_{0}\), initialize their Bezier-path control points \(\mathbf{P}\) and colors \(\mathbf{C}\).

\BlankLine
\textbf{Step 3: Optimize Paths at Coarsest Scale (Sec.~\ref{sec:redrawing})}\;
\For(\tcp*[h]{Gradient-descent loop}){$j = 1$ to $t_{\max}$}{
  Render image \(\hat{I}_0\) from \(\mathbf{P}, \mathbf{C}\). \\
  Compute MSE and Xing-loss \cite{xu2022live} (cf.\ Sec.~\ref{sec:iter_optim}). \\
  Update \(\mathbf{P}\) and \(\mathbf{C}\) using learning rates \(\alpha, \beta\). 
}
Update the pixel-wise loss weight based on the final MSE.

\BlankLine
\textbf{Step 4: Iterative Refinement Across Scales (Sec.~\ref{sec:iter_optim})}\;
\For(\tcp*[h]{From finer scales $I_{1}$ to $I_{T}$}){$t = 1$ to $T$}{
  Identify or refine new components at scale \(I_{t}\). \\
  Append new paths/colors \(\mathbf{P}, \mathbf{C}\) for these components. \\
  \For{$j = 1$ to $t_{\max}$}{
    Render \(\hat{I}_t\). \\
    Compute MSE and Xing-loss (Sec.~\ref{sec:iter_optim}). \\
    Update \(\mathbf{P}, \mathbf{C}\) via gradient descent. 
  }
  Update pixel-wise loss weight based on MSE at scale \(I_{t}\).
}

\BlankLine
\KwRet{Final Denoised Image \{\(\mathbf{P}, \mathbf{C}\)\} \tcp*[r]{Denoised representation}}
\end{algorithm}

\subsection{Step 4: Loss Functions}
A commonly used loss function to measure the discrepancy between the noisy image $I \in \mathbb{R}^{w \times h \times 3}$ and a rendered output 
\( \hat{I}_t \in \mathbb{R}^{w \times h \times 3} \)
is the \emph{mean square error (MSE)}. 
Despite its simplicity, MSE remains effective for many image comparison and reconstruction tasks because it is straightforward to implement and computationally efficient.
It is possible possible for some Bezier Paths to become
self-interacted during the course of optimization, leading to
detrimental artifacts and improper topology. 
To this end, we adopt the self interaction Xing-loss \cite{xu2022live} to mitigate this problem. 
Assuming all the b´ezier curves in our paper are third order, by analyzing a number of optimized shapes, the \cite{xu2022live} found that a self-interacted path always intersects the lines of its control points, and vice versa. 
This suggests that instead of optimizing the b´ezier
path, one potential solution would be adding a constraint
on the control points.
\section{Results}
\label{sec_results}
\emph{Dataset.} 
We use AxFLAIR brain MRI, Cor-PD knee MRI, FastMRI, and Modl \cite{aggarwal2021model} datasets to show the efficacy of our model. 
\begin{table*}[t!]
\centering
\caption{Quantitative comparison of DnCNN, N2N, N2void, Ne2Ne, Restoformer, LAN, Anisotropic, nonlocal, Beltrami, and our method. We report PSNR (dB) / SSIM. ``-'' indicates no result was provided.}
\label{tab:cross_domain_denoise}
\resizebox{\textwidth}{!}{
\begin{tabular}{ll|c|c|c|c|c}
\toprule
\multirow{2}{*}{\textbf{Model}} & \multirow{2}{*}{\textbf{Method}} & \multirow{2}{*}{$\mathbf{\sigma}$} 
& \multirow{2}{*}{\textbf{PSNR/SSIM}} & \multirow{2}{*}{\textbf{PSNR/SSIM}} 
& \multirow{2}{*}{\textbf{PSNR/SSIM}} & \multirow{2}{*}{\textbf{PSNR/SSIM}} \\ 
\cline{4-7}
 &  &  & \textbf{AxFLAIR} & \textbf{Cor-PD knee} & \textbf{FastMRI Breast} & \textbf{FastMRI Prostate} \\
\midrule
\textbf{Training} & DnCNN \cite{7839189} & - 
  & $38.10 / 0.952$ & - & $36.60 / 0.930$ & - \\

 & \multirow{3}{*}{N2N \cite{pmlr-v80-lehtinen18a}} 
   & $5$  & $38.07 / 0.951$ & $38.08 / 0.951$ & $36.60 / 0.929$ & $36.60 / 0.929$ \\
 &    & $10$ & $38.04 / 0.950$ & $38.06 / 0.951$ & $36.59 / 0.928$ & $36.60 / 0.928$ \\
 &    & $20$ & $37.99 / 0.949$ & $38.02 / 0.949$ & $36.56 / 0.925$ & $36.56 / 0.925$ \\

 & \multirow{3}{*}{N2void \cite{krull2019noise2void}}
   & $5$  & $37.93 / 0.948$ & $37.95 / 0.948$ & $36.48 / 0.923$ & $36.46 / 0.923$ \\
 &    & $10$ & $37.76 / 0.943$ & $37.83 / 0.945$ & $36.29 / 0.915$ & $36.28 / 0.915$ \\
 &    & $20$ & $37.47 / 0.935$ & $37.67 / 0.941$ & $35.95 / 0.902$ & $36.02 / 0.904$ \\

 & \multirow{3}{*}{Ne2Ne \cite{Huang_2021_CVPR}}
   & $5$  & $38.12 / 0.952$ & $38.13 / 0.952$ & $36.69 / 0.931$ & $36.70 / 0.931$ \\
 &    & $10$ & $38.13 / 0.952$ & $38.14 / 0.952$ & $36.75 / 0.931$ & $36.77 / 0.931$ \\
 &    & $20$ & $38.13 / 0.952$ & $38.14 / 0.952$ & $36.78 / 0.930$ & $36.81 / 0.930$ \\

 & \multirow{3}{*}{Restoformer \cite{zamir2022restormer}}
   & $5$  & $38.23 / 0.955$ & $38.23 / 0.956$ & $36.56 / 0.936$ & $36.54 / 0.935$ \\
 &    & $10$ & $38.23 / 0.955$ & $38.25 / 0.955$ & $36.66 / 0.934$ & $36.65 / 0.934$ \\
 &    & $20$ & $38.17 / 0.953$ & $38.20 / 0.954$ & $35.69 / 0.931$ & $36.68 / 0.930$ \\
 & \multirow{3}{*}{LAN \cite{10656535}}
   & $5$  & $38.22 / 0.954$ & $38.16 / 0.953$ & $36.73 / 0.934$ & $36.66 / 0.932$ \\
 &    & $10$ & $38.29 / 0.955$ & $38.22 / 0.954$ & $36.79 / 0.936$ & $36.71 / 0.933$ \\
 &    & $20$ & $38.29 / 0.955$ & $38.31 / 0.956$ & $36.78 / 0.938$ & $36.80 / 0.935$ \\
\midrule

\textbf{Untrained}  & \multirow{3}{*}{Anisotropic \cite{perona1990scale}}
   & $5$  & $39.09 / 0.966$ & $39.04 / 0.965$ & $38.14 / 0.952$ & $38.07 / 0.951$ \\
 &    & $10$ & $39.12 / 0.965$ & $39.04 / 0.965$ & $38.23 / 0.952$ & $38.08 / 0.950$ \\
 &    & $20$ & $39.14 / 0.965$ & $38.98 / 0.964$ & $38.35 / 0.953$ & $38.05 / 0.948$ \\

 & \multirow{3}{*}{nonlocal \cite{guidotti2009new}}
   & $5$  & $39.04 / 0.965$ & $39.00 / 0.965$ & $38.12 / 0.951$ & $38.07 / 0.950$ \\
 &    & $10$ & $38.96 / 0.964$ & $38.89 / 0.964$ & $38.05 / 0.950$ & $37.95 / 0.948$ \\
 &    & $20$ & $38.74 / 0.961$ & $38.66 / 0.961$ & $37.65 / 0.943$ & $37.52 / 0.941$ \\

 & \multirow{3}{*}{Beltrami \cite{ben2023denoising}}
   & $5$  & $39.07 / 0.965$ & $39.08 / 0.965$ & $38.09 / 0.951$ & $38.10 / 0.951$ \\
 &    & $10$ & $39.06 / 0.965$ & $39.07 / 0.965$ & $38.12 / 0.950$ & $38.14 / 0.950$ \\
 &    & $20$ & $39.02 / 0.964$ & $39.03 / 0.964$ & $38.12 / 0.948$ & $38.14 / 0.948$ \\
\midrule
 & \multirow{3}{*}{\textbf{Ours}}
   & $5$  & $39.23 / 0.968$ & $39.09 / 0.967$ & $38.31 / 0.957$ & $38.14 / 0.953$ \\
 &    & $10$ & $39.30 / 0.969$ & $39.14 / 0.967$ & $38.58 / 0.961$ & $38.25 / 0.955$ \\
 &    & $20$ & $39.28 / 0.969$ & $39.17 / 0.968$ & $38.86 / 0.965$ & $38.38 / 0.958$ \\
\bottomrule
\end{tabular}
} 
\end{table*}

\emph{Qualitative Comparison.} 
Table~\ref{tab:cross_domain_denoise} presents a comprehensive comparison of representative denoising approaches, including both supervised (e.g., DnCNN~\cite{7839189}, N2N~\cite{pmlr-v80-lehtinen18a}, Ne2Ne~\cite{Huang_2021_CVPR}) and unsupervised or self-supervised strategies (e.g., Aniso~\cite{perona1990scale}, nonlocal~\cite{guidotti2009new}, Beltrami~\cite{ben2023denoising}), as well as our proposed method. Performance is reported in terms of PSNR and SSIM under varying noise levels \(\sigma\) (\(5\), \(10\), and \(20\)), across four anatomical regions—Knee, Head, Breast, and Prostate. A dash (``-'') indicates that no result was available for that specific condition. Overall, the results demonstrate that our approach consistently outperforms most baseline methods on multiple tissue types, confirming its robustness and versatility in cross-domain medical image denoising.
\begin{figure}
    \centering
    \includegraphics[width=\linewidth]{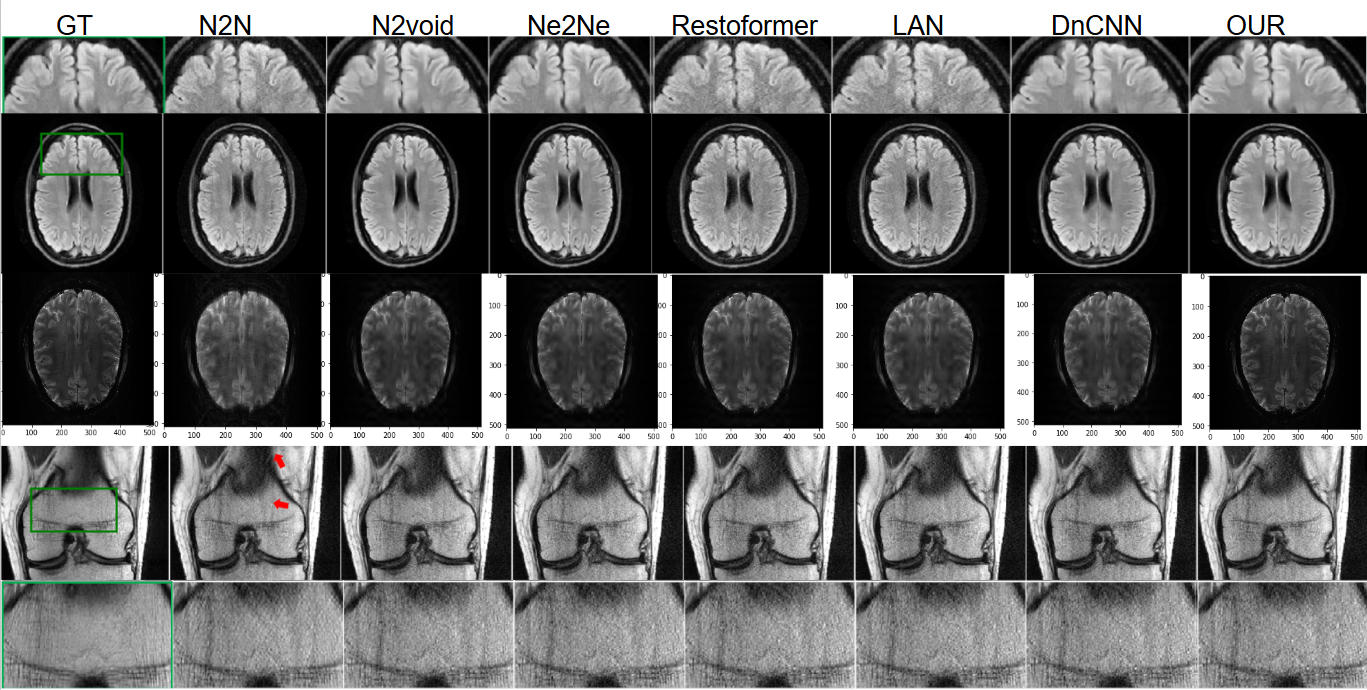}
    \caption{(Top) AxFLAIR brain MRI, (Bottom) Cor-PD knee MRI; qualitative comparison of DnCNN~\cite{7839189}, N2N~\cite{pmlr-v80-lehtinen18a}, Ne2Ne~\cite{Huang_2021_CVPR}, Restoformer \cite{zamir2022restormer}, N2void \cite{krull2019noise2void}, LAN \cite{10656535}, and our method. First column represent ground truth image.}
    \label{fig:11}
\end{figure}

\begin{figure}
    \centering
    \includegraphics[width=\linewidth]{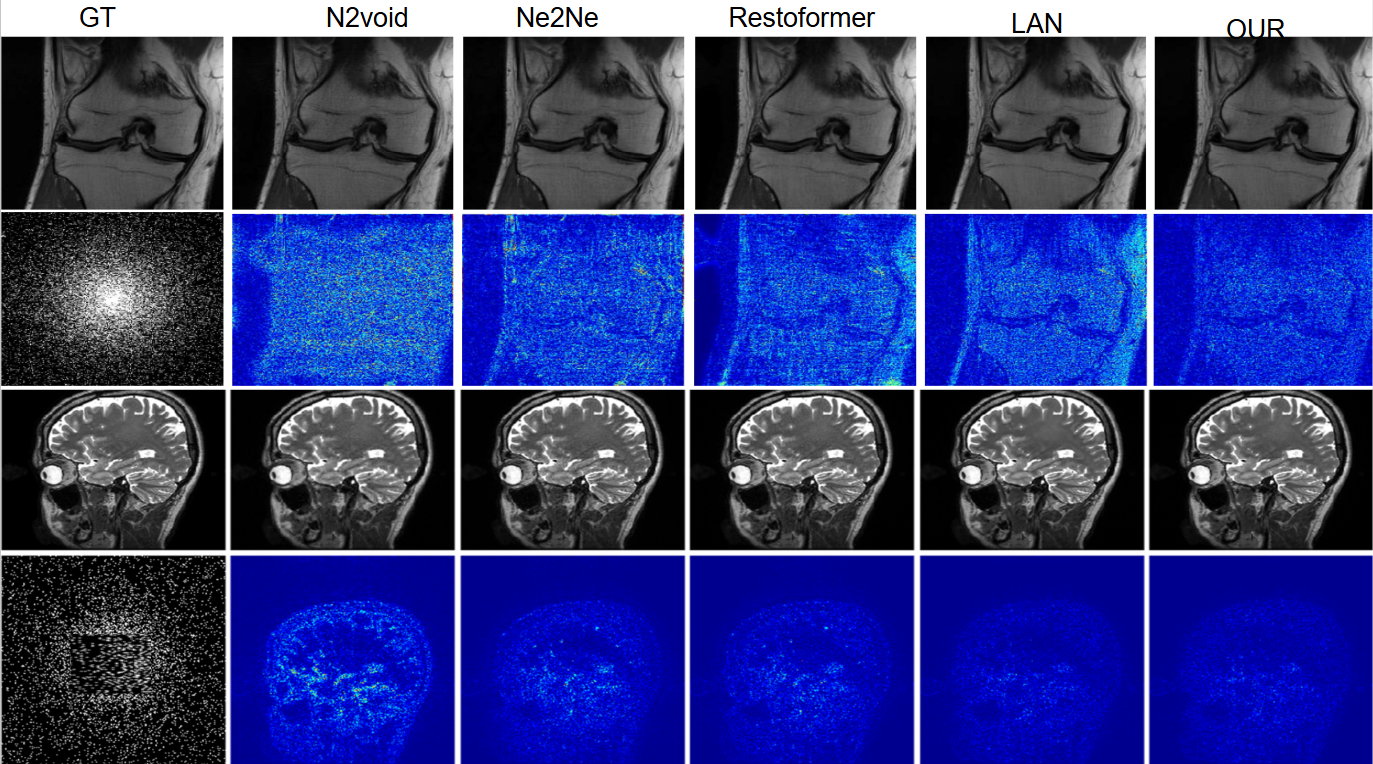}
    \caption{(Top) FastMRI knee reconstruction, (Bottom) Modl \cite{aggarwal2021model} Brain MRI; qualitative comparison of  Ne2Ne~\cite{Huang_2021_CVPR}, Restoformer \cite{zamir2022restormer}, N2void \cite{krull2019noise2void}, LAN \cite{10656535}, and our method. First column represent ground truth image.}
    \label{fig:1}
\end{figure}

\emph{Qualitative Comparisons.} 
We show in Figs \ref{fig:1} and \ref{fig:11} to show the qualitative comparison of our images. 
From the enlarged view of Fig. \ref{fig:11}, it is not difficult to find that the image reconstructed by DnCNN~\cite{7839189}, N2N~\cite{pmlr-v80-lehtinen18a}, Ne2Ne~\cite{Huang_2021_CVPR}, Restoformer \cite{zamir2022restormer}, N2void \cite{krull2019noise2void}, LAN \cite{10656535} methods. 
Fig. \ref{fig:11} shows the denoising results of trained models on a noise pattern $\sigma=15$, see first image of the second row in Fig. \ref{fig:11}. The experimental results are consistent with the above experimental performance, which further verifies the
robustness of our method in terms of noise pattern. 
\section{Discussion and Conclusion}
\label{sec:discussion_conclusion}

In this work, we introduced a multi-scale approach for denoising medical images by combining anisotropic Gaussian filtering and progressive Bezier-path redrawing. Through extensive evaluation on diverse MRI datasets (including AxFLAIR and Cor-PD knee MRI), our method demonstrates robust performance in terms of PSNR and SSIM, while preserving essential anatomical details such as tissue boundaries and subtle textural structures. A key advantage of our framework lies in its adaptability: by tuning the anisotropic blur parameters, our approach can address a wide range of noise levels without demanding extensive labeled training data.

Despite these benefits, there are some practical considerations. First, the iterative nature of the Bezier-path redrawing introduces a computational overhead, especially for high-resolution images. Approaches to accelerate convergence, such as leveraging multi-GPU or parallelized rendering strategies, could further improve usability in time-sensitive clinical settings. Second, while the anisotropic blur provides flexibility across different noise characteristics, further exploration of adaptive parameter selection (e.g., learning the blur hyperparameters) might enhance robustness in scenarios involving highly heterogeneous noise distributions.

In conclusion, our multi-scale denoising framework addresses both low- and high-frequency noise in a principled, coarse-to-fine manner. The proposed method effectively balances shape preservation and noise suppression, outperforming many conventional and learned denoising baselines across various anatomical regions. Future work will focus on extending our approach to 3D volumetric data and exploring how semantic information (e.g., segmentation masks) may be incorporated to guide the redrawing process. We believe that these developments will pave the way for more accurate and clinically valuable image reconstructions in numerous medical imaging applications.

\bibliographystyle{splncs04}
\bibliography{mybibliography}
\end{document}